\documentclass[aps,prl,reprint,superscriptaddress,letterpaper,twocolumn,longbibliography,floatfix]{revtex4-1}
\usepackage{graphicx}
\usepackage{dcolumn}   
\usepackage{bm}        
\usepackage{amssymb}   
\usepackage{amsmath}   
\usepackage{microtype} 
\usepackage{hyperref}  
\usepackage[cal=cm]{mathalpha}

\hypersetup{
  colorlinks=true,
  linkcolor=blue,
  citecolor=blue,
  urlcolor=blue,
  filecolor=blue
}

\DeclareSymbolFont{cmmath}{OMS}{cmsy}{m}{n}

\begin{document}


\title{Exact solution of the infinite-range dissipative transverse-field Ising model}

\author{David Roberts$^{1,2}$, A. A. Clerk}
\affiliation{Pritzker School of Molecular Engineering, University of Chicago, Chicago, IL, USA \\
$^2$Department of Physics, University of Chicago, Chicago, IL, USA}

\date{\today}

\begin{abstract} 
The dissipative variant of the Ising model in a transverse field is one of the most important models in the analysis of open quantum many-body systems, due to its paradigmatic character for understanding driven-dissipative quantum phase transitions, as well as its relevance in modelling diverse experimental platforms in atomic physics and quantum simulation. Here, we present an exact solution for the steady state of the transverse-field Ising model in the limit of infinite-range interactions, with local dissipation and inhomogeneous transverse fields. Our solution holds despite the lack of any collective spin symmetry or even permutation symmetry.  It allows us to investigate first- and second-order dissipative phase transitions, driven-dissipative criticality, and captures the emergence of a surprising ``spin blockade" phenomenon.  The ability of the solution to describe spatially-varying local fields provides a new tool to study disordered open quantum systems in regimes that would be extremely difficult to treat with numerical methods.  
\end{abstract}

\maketitle

{\it Introduction}. Dissipative transverse-field Ising (DTI) models are a paradigmatic class of open quantum systems in atomic physics, quantum information and quantum simulation. In these systems, a collection of spins interact on a lattice via Ising interactions, while subject to local magnetic fields that are transverse to the interaction axis; these fields can represent local Rabi drives in a rotating frame. Markovian dissipation is most commonly introduced via Lindblad dynamics with jump operators that induce either local dephasing \cite{marcuzziUniversalNonequilibriumProperties2014, borishTransverseFieldIsingDynamics2020} or local $T_1$ relaxation \cite{blochNuclear1946} along chosen axes \cite{jinPhaseDiagramDissipative2018,leeCollectiveQuantumJumps2012,kazemiDrivenDissipativeRydbergBlockade2023,singhDrivenDissipativeCriticalityDiscrete2022, weimerVariationalAnalysisDrivendissipative2015, overbeckMulticriticalBehaviorDissipative2017}. 
Interest in this model stems from its direct relevance in understanding diverse experimental platforms ranging across atomic physics. In dilute Rydberg gases, the mean-field equations of the DTI have been used to qualitatively capture bistability and hysteresis in the magnetization density when sweeping laser detunings \cite{carrNonequilibriumPhaseTransition2013,malossiFullCountingStatistics2014, paris-mandokiFreeSpaceQuantumElectrodynamics2017}. In quantum simulation platforms such as trapped ions, the DTI model has been realized with tunable power-law decaying interactions $J_{ij}= J_0/|i-j|^\alpha$ that approach the infinite-range limit with $0 <\alpha< 3$ \cite{brittonEngineeredTwodimensionalIsing2012a}. More recently, arrays of Rydberg atoms, with much shorter-range and stronger Ising interactions, have emerged as a versatile platform for simulating many-body quantum systems, and have realized the DTI on, e.g. triangular and hypercubic lattices \cite{ebadiQuantumPhasesMatter2021,schollQuantumSimulation2D2021}.

\begin{figure}
    \centering
    \includegraphics[width=0.8\columnwidth]{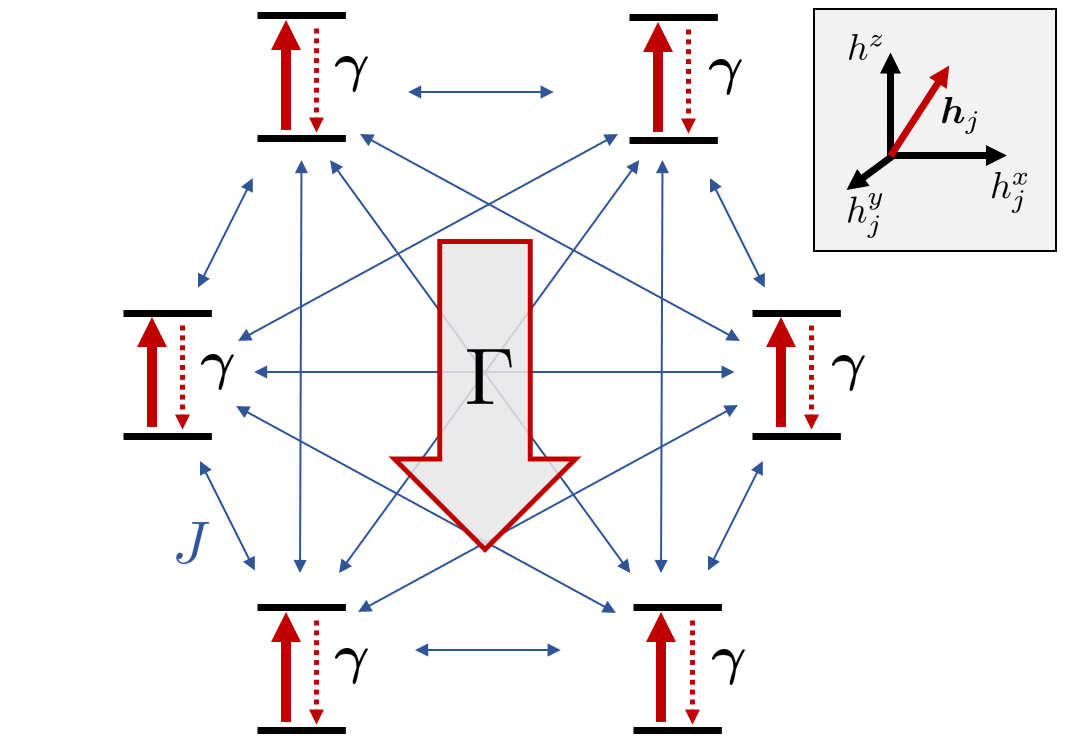}
    \caption{Schematic of our solvable dissipative transverse-field Ising model. A collection of $N$ spin-1/2 particles, represented here as a collection of driven-dissipative two-level systems, is subject to local $T_1$ decay at a rate $\gamma$, as well as collective decay at a rate $\Gamma$. The two-level systems interact via an all-to-all Ising interaction with strength $J$. Inset: each spin is subject to a separate external magnetic field with both axial and transverse components. The axial component of the external field is the same for each spin, whereas the transverse part can vary.}
    \label{fig:intro_fig}
\end{figure}

Solutions for the DTI nonequilibrium steady state have been lacking, mainly due to the difficulty in understanding the competition between the transverse field and the dissipation. Without a transverse field, the Lindblad dynamics is integrable due to the existence of a complete set of commuting weak symmetries \cite{foss-feigNonequilibriumDynamicsArbitraryrange2013,mcdonaldExactSolutionsInteracting2022}, and the steady-state problem is essentially a classical one. However, when the transverse field is present, even in the simplest case $\alpha = 0$, an analytical solution that is valid in all parameter regimes is still lacking. In this Letter, we address this issue by finding a ``hidden" symmetry \cite{robertsHiddenTimeReversalSymmetry2021,fagnolaGeneratorsKMSSymmetric2010} in the dynamics of the DTI model, leading to an exact solution for its dissipative steady state in the infinite-range limit. Our analytical solution remains valid for systems with inhomogeneous transverse fields, where the dynamics is not simplified by permutation symmetry.

Equipped with the exact solution, we are able to derive closed-form expressions for correlation functions of any order, offering valuable insights into the system's behavior. We also investigate a novel spin blockade effect characterized by unconventional correlation properties, and uncover an effective ``thermodynamic potential" which allows us to fully understand the large-$N$ limit, and which incorporates non-mean-field information, including the location of first-order phase transitions in regimes far from the critical point \cite{marcuzziUniversalNonequilibriumProperties2014,pazDrivendissipativeIsingModel2021}.

{\it Dissipative transverse-field Ising model}. We consider a dissipative system consisting of spin-$1/2$ particles described by the local spin observables $\hat{S}^\mu_j=\hat{\sigma}^\mu_j/2$, with $\mu = x,y,z$ denoting the spin direction and $j=1,2,\dots, N$ indexing the lattice sites. Here, $\hat{\sigma}^\mu_j$ denote the Pauli matrices. The spins are coupled via an all-to-all Ising interaction, and we allow for each lattice site to experience a different magnetic field $h_j^\mu$, which can have both transverse and axial components:
\begin{align}
\hat{H}_\text{TFIM} = J\sum_{i\neq j}\hat{S}^z_i\hat{S}^z_j +\sum_{j,\mu} h^\mu_j \hat{S}^\mu_j \label{eq:TFIMham}
\end{align}
This Hamiltonian can be viewed as representing a collection of Rabi-driven two-level systems in the rotating frame, with the transverse- and axial fields
corresponding to the drive amplitude and detuning of each local two-level system.

The full dynamics of our solvable model \cite{singhDrivenDissipativeCriticalityDiscrete2022, huybrechtsValidityMeanfieldTheory2020} also includes both collective and local $T_1$ decay for each two-level system and is described by the Lindblad master equation
\begin{align}
\displaystyle{\dot{\hat{\rho}} = -i[\hat{H}_\text{TFIM},\hat{\rho}] + \gamma\sum_j \mathcal D[\hat{\sigma}^-_j]\hat{\rho} + \Gamma\mathcal D[\hat{S}_-]\hat{\rho}}\equiv \mathcal L\hat{\rho},\label{eq:the_master_eqn}
\end{align}
where $\mathcal D[\hat{X}]\hat{\rho} \equiv \hat{X}\hat{\rho} \hat{X}^\dagger -(1/2) \{\hat{X}^\dagger \hat{X},\hat{\rho}\}$ is the usual Lindblad dissipative superoperator, $\gamma \equiv 1/T_1$ is the local longitudinal relaxation rate, and $\Gamma$ is a rate for collective relaxation that is relevant in many platforms,  capturing effects such as superradiant decay \cite{carrNonequilibriumPhaseTransition2013}. Finally, $\hat{S}_-:=\hat{S}^x - i\hat{S}^y$ is the lowering operator for the collective $SU(2)$ representation $\hat{S}^\mu :=\sum_j \hat{S}^\mu_j$. Throughout the rest of this Letter, we will fix the axial component of the magnetic field to be spatially uniform ($h^z_j \equiv h^z$), and will (without loss of any generality) set $h^y_j = 0$.

We focus exclusively on finding the steady states of this system, i.e. density matrices $\hat{\rho}_\text{SS}$ satisfying
\begin{align}
\mathcal L\hat{\rho}_\text{SS}=0.\label{eq:ss_eqn}
\end{align}
We briefly summarize prior work on this model. Even in the case that the external field $h^\mu_j$ is uniform, there has been no known exact solution for its steady state, although there has been much work studying asymptotic expansions in the semiclassical limit $N\to \infty$ \cite{leeCollectiveQuantumJumps2012,leeDissipativeTransversefieldIsing2013,atesDynamicalPhasesIntermittency2012,weimerVariationalAnalysisDrivendissipative2015, pazDrivendissipativeIsingModel2021}. Mean-field approximations for the steady states satisfying Eq. \eqref{eq:ss_eqn} have been used to qualitatively model dissipative phase transitions in dilute Rydberg gases trapped in, e.g. vapor cells \cite{carrNonequilibriumPhaseTransition2013,malossiFullCountingStatistics2014,marcuzziUniversalNonequilibriumProperties2014}, and have been proven to be exact in the limit $N\to\infty$ \cite{carolloExactness2021, fiorelliMeanfieldDynamicsOpen2023a}. For $N$ finite, the dissipative steady state has been studied using numerical methods that take advantage of the permutation symmetry of the problem \cite{huybrechtsMeanfieldValidityDissipative2020,joResolvingMeanfieldSolutions2022}. \\


{\it Exact steady-state solution}.  Suppose that $\mathcal L$ were the Liouvillian for a classical master equation. In this case, a common method for solving Eq. \eqref{eq:ss_eqn} would be to impose detailed balance conditions on the steady-state probability density \cite{kelleyf.p.ReversibilityStochasticNetworks1979}. In our case, we factor the steady state into two pieces, $\hat{\rho}_\text{SS}=\hat{\Psi}\hat{\Psi}^\dagger$, with $\hat{\Psi}$ antilinear, and impose the {\it quantum} detailed balance conditions \cite{fagnolaGeneratorsDetailedBalance2007,fagnolaGeneratorsKMSSymmetric2010}
\begin{align}
    \hat{H}_\text{eff}\hat{\Psi} = \hat{\Psi}\hat{H}_\text{eff}^\dagger,~~~~~\hat{L}_k\hat{\Psi} = \hat{\Psi}\hat{L}^\dagger_k,\label{eq:detailedbalance}
\end{align}
where $\hat{L}_k\in \{\hat{\sigma}^-_1,\dots, \hat{\sigma}_N^-, \hat{S}_-\}$ ranges over the set of Lindblad operators in (2), and $\hat{H}_\text{eff}\equiv \hat{H}-(i/2)\sum_k \hat{L}_k^\dagger\hat{L}_k$ is an effective non-Hermitian Hamiltonian which allows us to write the Lindbladian in the form 
\begin{align}
    \mathcal L\hat{\rho} &=-i(\hat{H}_\text{eff}\hat{\rho} - \hat{\rho} \hat{H}_\text{eff}^\dagger)+ \sum_k \hat{L}_k \hat{\rho} \hat{L}_k^\dagger.\label{eq:the_lindbladian_Heff_form}
\end{align}
Using \eqref{eq:the_lindbladian_Heff_form}, we confirm that if $\hat{\Psi}$ is a solution to the detailed balance conditions \eqref{eq:detailedbalance}, then $\hat{\rho}:=\hat{\Psi}\hat{\Psi}^\dagger$ is a valid steady state of the master equation. 

To solve Eqs.~\eqref{eq:detailedbalance}, we write $\hat{\Psi}=\hat{\Phi}\hat{K}$, with $\hat{K}=\hat{K}_z\hat{U}_x$, where $\hat{K}_z$ denotes complex conjugation in the eigenbasis of the commuting operators $\hat{\sigma}^z_1,\dots, \hat{\sigma}^z_N$, and $\hat{U}_x:=\prod_j \hat{\sigma}^x_j$ is the 
global $x$-parity operator. Remarkably, whenever $\Gamma \equiv 0$ or the transverse fields are uniform, i.e. $h^{x}_j \equiv h^{x}$, we can explicitly solve for $\hat{\Phi}$, and hence obtain $\hat{\rho}_\text{SS}$. In both situations, we obtain \cite{supp}
\begin{align}
    \hat{\rho}_\text{SS}=\hat{\Phi}\hat{\Phi}^\dagger/\mathcal N,~~~~~~~~\hat{\Phi}=(1-\hat{\mathbb{S}}_-)^{h^z_\text{eff}/J_\text{eff}},\label{eq:the_soln}
\end{align}
where we have defined an effective complex Ising coupling $J_\text{eff}:=J+i\Gamma/2$, and an effective complex longitudinal field $h^z_\text{eff}:=h^z-i(\gamma + \Gamma)/2$.  Finally, $\mathcal N\equiv \text{tr}(\hat{\Phi}\hat{\Phi}^\dagger)$ is a normalization constant. We have also defined a non-unitary representation of $SU(2)$,
\begin{align}
    \hat{\mathbb{S}}_- :=\sum_j\frac{2J_\text{eff}\hat{\sigma}^-_j}{h_j^x},~~~~\hat{\mathbb{S}}_+:=\sum_j \frac{h_j^x\hat{\sigma}^+_j}{2J_\text{eff}}. \label{eq:su2}
\end{align}
Explicitly, $\hat{\mathbb{S}}^x := (\hat{\mathbb{S}}_++\hat{\mathbb{S}}_-)/2$, $\hat{\mathbb{S}}^y:=(\hat{\mathbb{S}}_+-\hat{\mathbb{S}}_-)/2i$, along with $\hat{\mathbb{S}}^z :=\hat{S}^z$ satisfy the commutation relations of the $SU(2)$ algebra. However, since $\hat{\mathbb{S}}^\dagger_- \neq \hat{\mathbb{S}}_+$, $\hat{\rho}_\text{SS}$ does not in general preserve the decomposition of the Hilbert space into irreducible representations for the $\hat{\mathbb{S}}^\mu$ operators, reflecting the non-collective nature of the dissipative dynamics. In the fully collective limit $\gamma \equiv 0$ with no Ising interaction or inhomogeneity, we recover the solution of the driven Dicke model \cite{puriExactSteadystateDensity1979,drummondObservablesMomentsCooperative1980,lawandeNonresonantEffectsFluorescent1981, schneiderEntanglementSteadyState2002, nagyDickeModelPhaseTransition2010, hannukainenDissipation2018}. In what follows, we investigate the physics that emerges from our solution (\ref{eq:the_soln}, \ref{eq:su2}).

\begin{figure}
    \centering
    \includegraphics[width=0.99\columnwidth]{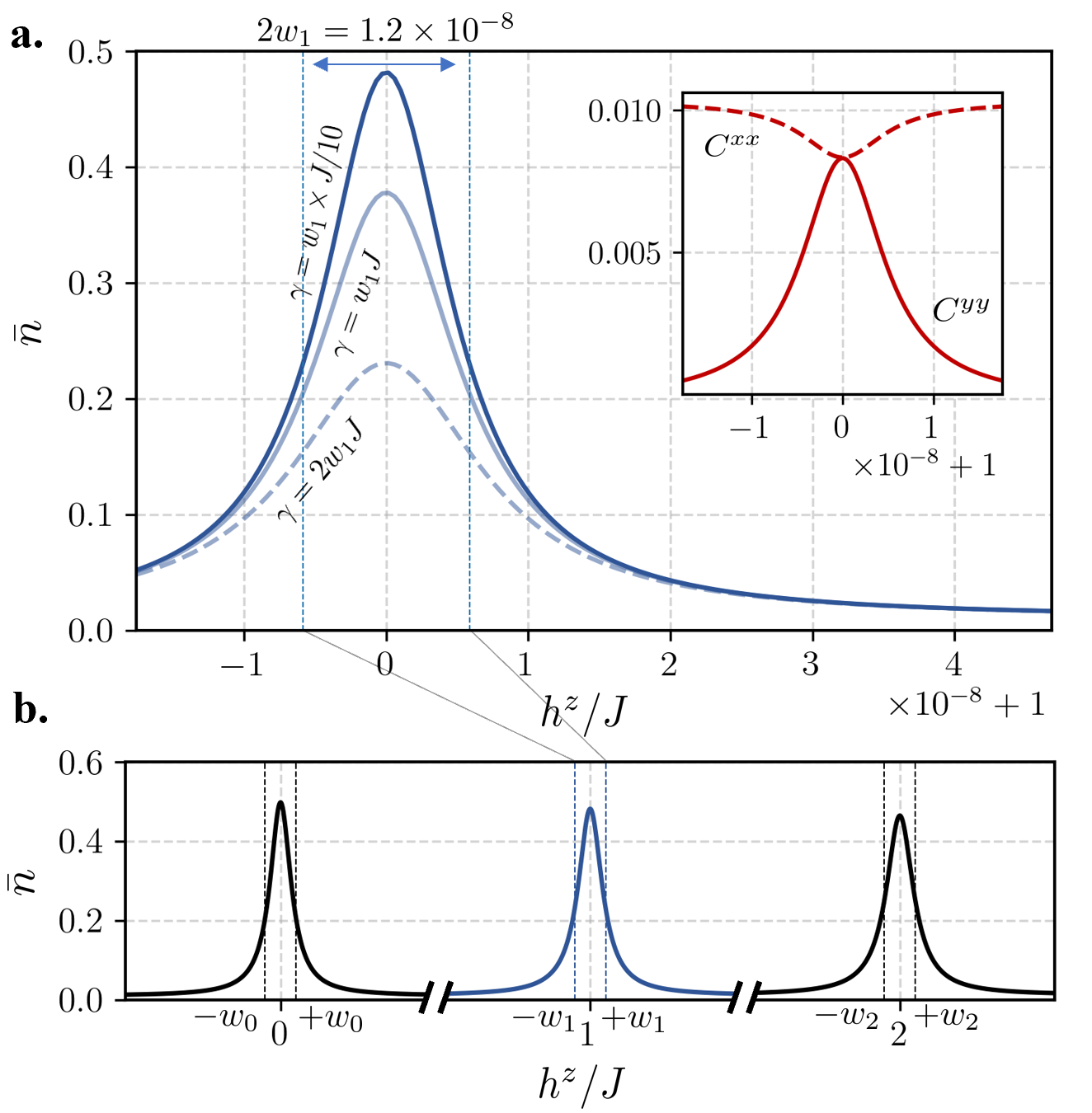}
    \caption{Blockade effect in the dissipative transverse-field Ising model. 
    Top panel (a): plot of the density of up-spins, $\bar{n}=\sum_j\langle \hat{\sigma}^+_j\hat{\sigma}^-_j\rangle_\text{SS}/N$, near a first-order blockade, for $h^x=4J$, $\Gamma = 0$, and $N=22$. Inset: Plot of the correlation functions $C^{xx}$ (dashed red line) and $C^{yy}$ (solid red line) for the same parameters, where $C^{\mu\mu} := \langle \hat{\sigma}^\mu_1\hat{\sigma}^\mu_2\rangle_\text{SS}$. Bottom panel (c): Close-ups of the first three blockades, $n=0,1,2$. Parameters are the same as in panel (b). The exact solution yields extremely precise estimates for the widths of these features, see \eqref{eq:blockade_estimate} for more details.}
    \label{fig:blockade_figure}
\end{figure}

{\it Spin blockade}. Setting $h^z_\text{eff}/J_\text{eff}=0,1,2,\cdots$ yields an interesting class of states:
\begin{align}
    \hat{\Phi} &= (1-\hat{\mathbb{S}}_-)^n,~~~~~n=0,1,2,\dots~~~\label{eq:blockadedstate} 
\end{align}
This condition can be realized by setting $h^z = nJ$, where $n$ is a non-negative integer, and letting $\gamma,\Gamma\to 0^+$. The condition on the longitudinal field can be understood as a resonance between all spins down and a configuration where all but $n$ spins are excited. In this situation, even for drive amplitudes $h^x \lesssim J$, which would seem to preclude being able to excite many spins starting from an all-down state, one obtains a steady state with an extremely high excitation density, see Figure \ref{fig:blockade_figure}.
The truncated form of the steady state at a blockade value of $h^z$ leads to extremely subtle correlation properties. In particular, the rescaled coherence
\begin{align}
    h_{i_1}^x\cdots h_{i_n}^x\langle \hat{\sigma}^+_{i_1}\cdots \hat{\sigma}^+_{i_n}\rangle_\text{SS}
\end{align}
is independent of $i_1,\cdots, i_n$. Furthermore, the $k$th-order coherence for $k>n$ vanishes. In the simplest case $n=0$, the steady state is a completely depolarized state, and {\it all} coherences vanish. The case $n=1$ is nontrivial. In this situation, the rescaled transverse magnetization $h_j^x\langle \hat{\sigma}_j^+\rangle_\text{SS}$ is independent of $j$. Since 
\begin{align}
    \text{Re}\,\langle \hat{\sigma}^+_i\hat{\sigma}^+_j\rangle_\text{SS} =  \langle \hat{\sigma}^x_i\hat{\sigma}^x_j\rangle_\text{SS} - \langle \hat{\sigma}^y_i\hat{\sigma}^y_j\rangle_\text{SS},    
\end{align}
the transverse correlation functions coincide in this limit, c.f. the inset in Figure \ref{fig:blockade_figure}a.

If we also set $|J_\text{eff}/h^x_j| \gg 1$, then we can further approximate $\hat{\Phi}\approx \hat{\mathbb{S}}_-^n$. In this limit, there is a vanishing probability to observe a spin configuration with fewer than $n$ spins down in the steady state. In particular, the limiting distribution is a binomial distribution  \cite{supp}:
\begin{align}
    \text{Prob}(M^z = N-k) \sim \binom{N}{k}\binom{k}{n},
\end{align}
where here, $\hat{M}^\mu = 2\hat{S}^\mu$ is the total magnetization.

The exact solution allows us to obtain a precise bound on the modulus of $\epsilon\equiv h^z_\text{eff}/J_\text{eff} - n$ within which the above effects are observable. To do this, we consider the corrections to \eqref{eq:blockadedstate} by retaining the remaining terms in the binomial series. Assuming the transverse fields are uniform, a simple estimate comparing the Hilbert-Schmidt norms of the $n$th and $N$th terms in the series yields the bound $|\epsilon| < w_n$, with
\begin{align}
    w_n = \frac{1}{(N-n-1)!}\sqrt{\lambda^{n-N}\binom{N}{n}},\label{eq:blockade_estimate}
\end{align}
where we have defined $\lambda := 2|J_\text{eff}/h^x|^2$. The above bound yields extremely accurate estimates for the full-width at half maximum (FWHM) of each blockade feature, c.f. Figure \ref{fig:blockade_figure}b.

{\it Phase transitions and the large-$N$ limit}. At the mean-field level, as one sweeps the longitudinal field $h^z$ (corresponding to sweeping the drive-detuning in the driven qubit realization), a spin-configuration with $O(N)$ spins pointing up can become self-consistently resonant, leading to features of a first-order phase transition, such as bistability and hysteresis in the magnetization density \cite{marcuzziUniversalNonequilibriumProperties2014,carrNonequilibriumPhaseTransition2013}. Our exact solution yields several new insights into this transition, including the fact that this transition is fundamentally a transition between a polarized and a depolarized state, with the magnetization density serving as a {\it proxy} for the amount of polarization in the steady state. The exact solution also allows us to transparently see such a phase transition emerge in the large-$N$ limit, and reveals insights that go beyond mean-field theory, including the location of first-order phase transitions, as well as the behavior of multispin correlations across the transition.

Our starting point is the binomial expansion of $\hat{\Phi}$. Defining $r=h^z_\text{eff}/J_\text{eff}$, this is
\begin{align}
{\hat{\Phi}=}1-r\sum_{k=1}^N(1-r)\cdots (k-1-r)\frac{\hat{\mathbb{S}}_-^k}{k!}.
\end{align}
The relative weight of each monomial $\hat{\mathbb{S}}_-^m$ in this expansion determines an important piece of physics: if the lowest powers of $\hat{\mathbb{S}}_-$ dominate, $\hat{\rho}_\text{SS}$ is close to a maximally mixed (depolarized) state, and the magnetization density $m^\mu = \langle \hat{M}^\mu\rangle/N$ approaches zero. On the other hand, if the highest powers of $\hat{\mathbb{S}}_-$ dominate, the steady state is close to a completely polarized state and the (longitudinal) magnetization approaches its minimum value $m^z\equiv -1$.

We can make this intuition more precise. We start with the identity $\text{tr}(\hat{\mathbb{S}}_-^{\dagger M} \hat{M}^z \hat{\mathbb{S}}_-^{M'}) = -M\text{tr}(\hat{\mathbb{S}}_-^{\dagger M}\hat{\mathbb{S}}_-^{M'})$ involving the (longitudinal) magnetization. As a result, we have the somewhat suggestive formula 
\begin{align}
    \langle \hat{M}^z\rangle_\text{SS} = -\frac{\sum_M Mp_M}{\sum_M p_M}
    \label{eq:MzAverage}
\end{align}
where $p_M = \|\binom{r}{M}\hat{\mathbb{S}}_-^M\|^2_\text{HS}$ is the size of the term proportional to $\hat{\mathbb{S}}_-^M$ in the expansion of $\hat{\Phi}$ with respect to the Hilbert-Schmidt norm. Eq.~(\ref{eq:MzAverage}) makes precise the intuition described above: if the distribution $p_M$ is biased towards terms with $M \ll N$, then the
magnetization density vanishes in the thermodynamic limit, and the steady state is close to a depolarized state. The opposite occurs when the distribution is skewed towards terms with $M\to N$, in which case the steady state approaches a completely polarized state.

To understand the large-$N$ behavior, we write the magnetization density in the suggestive form
\begin{align}
    m^z = -\frac{\sum_m m\, e^{-Nf(m)}}{\sum_m \, e^{-Nf(m)}},    \label{eq:free_energy}
\end{align}
with $m=M/N$ now a variable ranging between zero and one, representing how polarized the steady state is. Here, $f(m)\equiv -N^{-1}\log p_{Nm}$ is a dimensionless ``free energy" that controls the relative weight of the terms in the expansion of $\hat{\Phi}$. To approach the thermodynamic limit, we must hold $\bar{J}_\text{eff}:= NJ_\text{eff}$ fixed, which fixes the renormalized parameters $\bar{r}:= h^z_\text{eff}/\bar{J}_\text{eff}$, and $\bar{\lambda}_j :=  2|\bar{J}_\text{eff}/h_j^x|^2$. 

To make it easier to calculate the free energy, we assume that the transverse fields are uniform, $h_j^x\equiv h^x$. In this case, we have an asymptotic series for $f(m)$ in the large-$N$ limit, with a leading-order contribution
\begin{align}
    f(m) &= \log \frac{m^m(1-m)^{1-m}}{(m-\bar{r})^{m-\bar{r}}(m-\bar{r}^*)^{m-\bar{r}^*}}-(\log\bar{\lambda}-2)m,
\end{align}
with $O(N^{-1})$ corrections to $f$ given in \cite{supp}. Since $r$ (and hence $\bar{r}$) cannot be a non-negative real number, $f$ must be an analytic function of $m$. As such, the analytic function $f(m)$ plays an analogous role to an equlibrium free energy in Landau theory \cite{landauTheoryPhaseTransitions1937}, and lets us rigorously describe the phase transition physics in our model. 

\begin{figure}
    \centering
    \includegraphics[width=1.00\columnwidth]{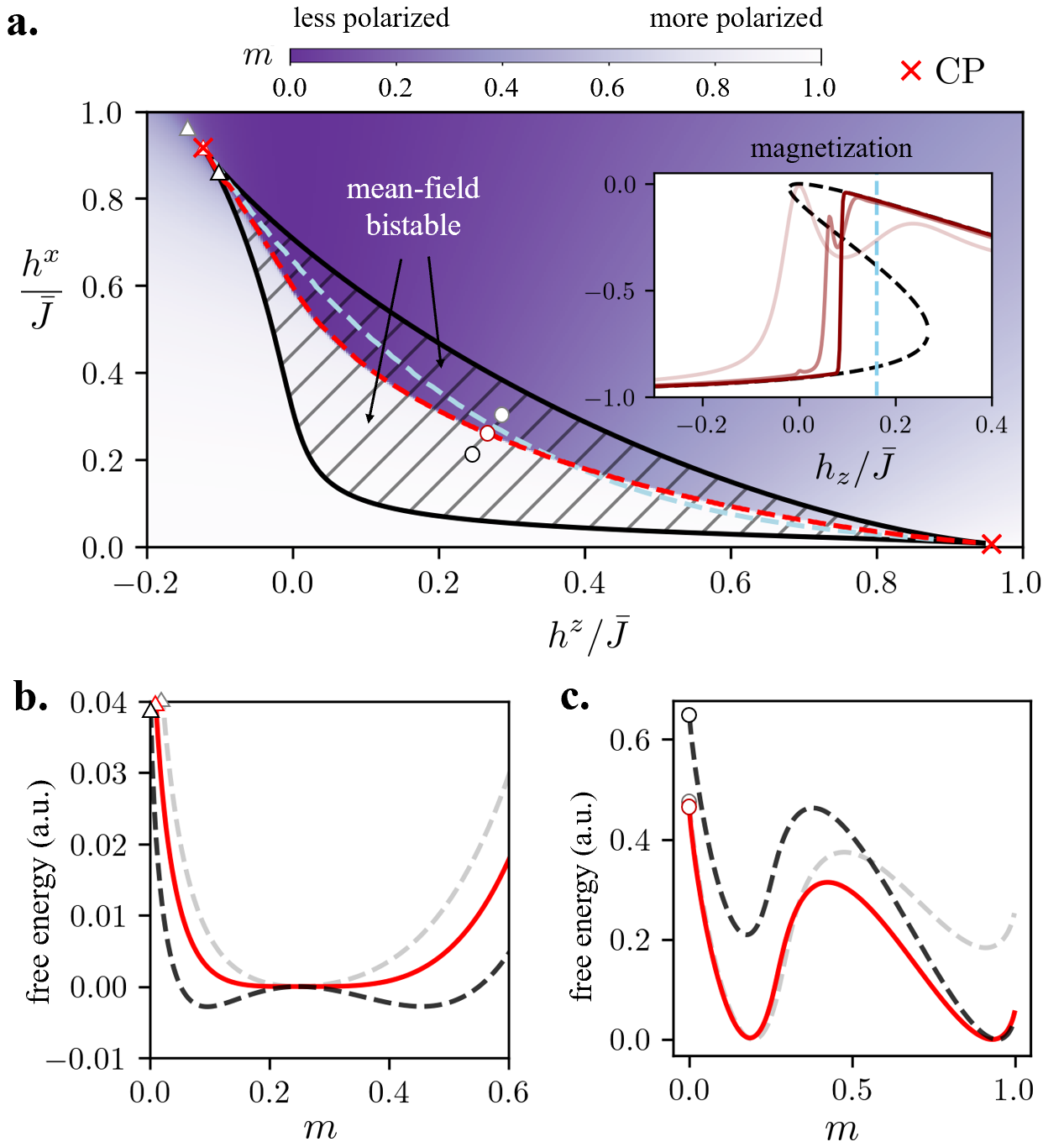}
    \caption{
    First- and second-order phase transitions in the dissipative transverse-field Ising model. Top panel (a): Steady-state polarization density $m=-\sum_j \langle \hat{\sigma}^z_j\rangle_\text{SS}/N$, for $\gamma = \bar{J}/10$, $N=128$, and $\Gamma = 0$. The shaded region enclosed by solid black lines denotes 
    parameters where mean field theory predicts bistability. The first-order phase transition occurs when the local minima in $f$ become degenerate, and is traced by a red dashed curve. The location of this transition predicted by the Maxwell construction is given by the  blue dashed curve. Critical points where the first three derivatives of $f$ vanish are marked by red crosses. Inset: Steady-state magnetization for $h^x = 0.4\bar{J}$, and $\gamma = \bar{J}/100$, for $N=4,16,128$. Increasing values of $N$ correspond to increasing opacity. Black dashed curves are used to denote the mean-field solutions. Bottom panel: free energy landscape $f(m)$ determining the amount of polarization $m$ in the steady state $\hat{\rho}_\text{SS}$, for three parameter choices near (b) a second-order phase transition, and (c) a first-order phase transition.
    }
    \label{fig:phase-transition-panel}
\end{figure}

In particular, the global minima $m^*$ of the potential $f$ correspond to the thermodynamic limit $\lim_{N\to\infty} m^z$ of the magnetization density. Whenever $f$ has {\it two} degenerate global minima, the magnetization density is the average of the two minima, and we have a first-order phase transition from a polarized to a depolarized state, c.f. Fig.~\ref{fig:phase-transition-panel}c. More details can be seen in Fig.~\ref{fig:phase-transition-panel}a, where we also show that the location of the first-order transition line in our model is not accurately predicted by a simple Maxwell construction (indicating the non-equilibrium nature of the model). Whenever the first three derivatives of $f$ vanish at a global minimum, we have a quantum critical point and a second-order phase transition, c.f. Fig.~\ref{fig:phase-transition-panel}b. Note that the formula \eqref{eq:free_energy} is still valid for the case of an inhomogeneous magnetic field. In \cite{supp}, we study these phase transitions in the presence of transverse-field disorder, and also connect them to a sign-change in a particular connected correlation function.

{\it Correlation functions}. Finally, the exact solution allows one to analytically solve for equal-time correlation functions of arbitrary order in our spin model. For the purpose of succinctly stating the main results, it is useful to define $\hat{\sigma}^\pm_{C }\equiv \prod_{j\in C}\hat{\sigma}^\pm_j$, for $C\subset \{1,2,3,\dots, N\}$ an arbitrary cluster of spins. Given such a cluster, it is convenient to use the notation $C^c\equiv \{1,2,3,\dots, N\}\setminus C$ to denote its complement, and $|C|$ to denote its size. Given a pair of clusters $C_1,C_2$, our exact solution yields the expectation value of the normally-ordered product $\hat{\sigma}^+_{C_1}\hat{\sigma}^-_{C_2}$. To express the answer in a compact form, we define auxiliary regions $R_{12} = C_1\cup C_2$, and $R_j \equiv (C_1\cup C_2)\setminus C_j$. In the situation where the external fields $h^\mu$ are uniform, one can express the solution for the expectation value in terms of generalized hypergeometric polynomials: 
\begin{align}
    \langle& \hat{\sigma}^+_{C_1}\hat{\sigma}^-_{C_2}\rangle_\text{SS}=\frac{2^{|R_{12}^c|}\mathcal{K}^*(R_1)\mathcal{K}(R_2)}{\mathcal N}\label{eq:homogeneous_correlations}\\
    &\times \,_3F_0(|R_1|-r^*,|R_2|-r, |R_{12}| - N;-2|J_\text{eff}/h^x|^2),\nonumber
\end{align}
where $\mathcal N = 2^N\,_3F_0(-r^*,-r, -N;-2|J_\text{eff}/h^x|^2)$. We have also defined form factors $\mathcal{K}(X) = (-r)_{|X|}\prod_{j\in X}(2J_\text{eff}/h_j^x)$, where $(z)_m \equiv z(z+1)\cdots (z+m-1)$ is the Pochhammer symbol, or rising factorial \footnote{In this case, the external fields are uniform, and so we can use the simpler formula $\mathcal{K}(X) = (-r)_{|X|}(2J_\text{eff}/h^x)^{|X|}$.}. 

More generally, in the case of an inhomogeneous field, we obtain the expression
\begin{align}
    \langle \hat{\sigma}^+_{C_1}\hat{\sigma}^-_{C_2}\rangle_\text{SS}&=\frac{2^{|R_{12}^c|}\mathcal{K}^*(R_1)\mathcal{K}(R_2)}{\mathcal N}\label{eq:inhomogeneous_correlations}\\
    &\times \sum_m(|R_1|-r^*)_m(|R_2| - r)_m\mathcal{I}_m(R_{12}^c),\nonumber
\end{align}
where we have defined the form factors $\mathcal{I}_m(X) \equiv \sum_{Y\subseteq X,\,|Y|=m}\,\prod_{j\in Y}(2|J_\text{eff}/h_j^x|^2)$. These form factors satisfy the recursion relation $\mathcal{I}_m(X) = \mathcal{I}_{m}(X\setminus \{j\}) + 2\frac{|J_\text{eff}|^2}{|h_j^x|^2}\mathcal{I}_{m-1}(X\setminus \{j\})$, which is useful when evaluating the expectation value in the thermodynamic limit.

{\it Summary \& Outlook}. Here, we present an exact solution for the steady state of a dissipative variant of the infinite-range transverse-field Ising model, including regimes where there is no collective structure or permutation symmetry. We uncover a novel spin blockade effect, and use an effective thermodynamic potential to analytically capture large-$N$ features of the DTI model.  While we have focused primarily on the fully-connected transverse-field Ising model, further studies could investigate the steady states of driven-dissipative spin systems defined on graphs with more nontrivial connectivity. In particular, the steady state of the DTI on a $d$-dimensional cubic lattice with $d<\infty$ remains an open problem, and it is interesting to see whether any of the physics we have uncovered in the fully-connected model is also applicable in this more nontrivial regime.

We would like to thank Martin Koppenhoefer and Paul Wiegmann for helpful discussions. This work was supported by the Air Force Office of Scientific Research MURI program under Grant No.~FA9550-19-1-0399, and the Simons Foundation through a Simons Investigator award (Grant No. 669487).

\bibliographystyle{apsrev4-1}
\bibliography{ms}

\end{document}


\selectlanguage{english}

\title{Supplementary Information for "Exact solution of the infinite-range dissipative transverse-field Ising model"}

\author{David Roberts$^{1,2}$, A. A. Clerk}
\affiliation{Pritzker School of Molecular Engineering, University of Chicago, Chicago, IL, USA \\
$^2$Department of Physics, University of Chicago, Chicago, IL, USA}
\date{\today}

\maketitle

\onecolumngrid


\section{Exact solution for the steady state}

The mixed-field Ising model, in the infinite-range limit, has the following Hamiltonian:
\begin{align}
    \hat{H}=J\sum_{i\neq j} \hat{S}^z_i\hat{S}^z_j + \sum_{j,\mu} h^\mu_j\hat{S}^\mu_j\label{eq:theham}
\end{align}
Here, $\mu=x,y,z$, and $\hat{S}^\mu_j\equiv \hat{\sigma}^\mu_j/2$ are the local $SU(2)$ operators used in the main text. We consider a Lindblad master equation with both collective and local loss, $\partial_t\hat{\rho} = -i[\hat{H},\hat{\rho}] + \mathcal{L}_1[\hat{\rho}] + \mathcal{L}_2[\hat{\rho}]$, that is, with
\begin{align}
    \mathcal L_1[\hat{\rho}] &=\frac{\gamma}{2}\sum_j \big(2\hat{\sigma}^-_j\hat{\rho} \hat{\sigma}^+_j - \{\hat{\sigma}^+_j\hat{\sigma}^-_j,\hat{\rho}\}\big),\\
    \mathcal L_2[\hat{\rho}] &=\frac{\Gamma}{2}\sum_j \big(2\hat{S}_-\hat{\rho} \hat{S}_+ - \{\hat{S}_+\hat{S}_-,\hat{\rho}\}\big).\label{eq:collectiveLossTerm}
\end{align}
The loss operators above, together with their adjoints, generate the entire algebra of observables for our spin system. Therefore,  thanks to the centralizer condition in \cite{evansIrreducible1977}, we know that the above Lindblad master equation must have a unique steady state. We now make the ansatz that this steady state satisfies a quantum detailed balance condition, namely $s$-QDB with $s=1/2$ \cite{goldsteinKMS1995, fagnolaGeneratorsKMSSymmetric2010, ramezaniQuantum2018}. A sufficient condition for $1/2$-QDB is that the steady state admits an antilinear square root $\hat{\Psi}$ with unit Hilbert-Schmidt norm, satisfying \cite{fagnolaGenerators2007}
\begin{align}
    \hat{O}\hat{\Psi} = \hat{\Psi} \hat{O}^\dagger,\label{eq:the_constraint_Psi}
\end{align}
for all linear operators $\hat{O}\in \text{span}\{H_\text{eff}, \hat{L}_1, \cdots, \hat{L}_l\}$, where $l=N+1$ is the number of jump operators. Letting $\hat{\Phi}:= \hat{\Psi}\hat{K}$ with $\hat{K}$ an antiunitary operator, we can gauge transform these conditions into
\begin{align}
    \hat{O}\hat{\Phi}=\hat{\Phi} (\hat{K}^\dagger \hat{O} \hat{K})^\dagger,~~~~~~~\forall \hat{O}\in \text{span}\{\hat{H}_\text{eff}, \hat{L}_1, \cdots, \hat{L}_l\}\label{eq:the_constraint_Phi}
\end{align}

\subsection{Connection to hidden TRS}
Before we proceed, it is perhaps useful to first review the equivalence between the quantum detailed balance conditions \cite{goldsteinKMS1995,fagnolaGenerators2007} used in this work and the "hidden" time-reversal symmetry conditions investigated in \cite{roberts_hidden_2020}. In particular, recall the hidden TRS conditions as formulated in Eq. (43) of \cite{roberts_hidden_2020}, but adapted to the case of the dissipative TFIM (without the collective loss, for simplicity). In particular, these conditions state that an antiunitary $\hat{T}$ is a hidden TRS if
\begin{align}
    \mathcal J[\hat{H}_\text{eff}] = \hat{H}_\text{eff}+E,~~~~\mathcal J[\hat{\sigma}_j^-] = \sum_{k=1}^{N} U_{jk} \hat{\sigma}_k^-,~~~~U^2 = 1,\label{eq:hTRS}
\end{align}
where $E$ is a real number, and $U_{jk}$ is an $N\times N$ unitary matrix, and where the exchange superoperator $\mathcal J$ is defined in Eq. (27) of \cite{roberts_hidden_2020} as $\mathcal J[\hat{X}] = \hat{\rho}^{1/2} \hat{T}\hat{X}^\dagger \hat{T}^{\dagger}\hat{\rho}^{-1/2}$, where $\hat{\rho}$ is the steady-state density matrix (which in the setting of \cite{roberts_hidden_2020} is assumed to be full rank). Writing $\hat{\Psi}:=\hat{\rho}^{1/2} \hat{T}$, we notice that the the exchange superoperator becomes
\begin{align}
    \mathcal J[\hat{X}] = \hat{\Psi}\hat{X}^\dagger\hat{\Psi}^{-1}
\end{align}
and furthermore $\hat{\Psi} \hat{\Psi}^\dagger = \hat{\rho}$, so that $\hat{\Psi}$ is an antilinear square root of the steady-state density matrix.\\

The above conditions are therefore equivalent to postulating that the steady state admits {\it some} antilinear square root $\hat{\Psi}$ such that the following quantum detailed balance condition \cite{fagnolaGenerators2007} holds:
\begin{align}
    (\hat{H}_\text{eff} + E)\hat{\Psi}&=  \hat{\Psi}\hat{H}_\text{eff}^\dagger,~~~~\sum_{k=1}^NU_{jk}\hat{\sigma}^-_k\hat{\Psi} = \hat{\Psi}\hat{\sigma}^+_j,~~~~U^2=1,\label{eq:generaldbcondition}
\end{align}
where $E$ is a real number, and $U_{jk}$ is an $N\times N$ unitary matrix. We recover the condition \eqref{eq:the_constraint_Psi} used in the main text by setting $U =1$ and $E=0$. We then recover the hidden time-reversal symmetry $\hat{T}$ as the antiunitary part of the polar decomposition of $\hat{\Psi}$.

\begin{figure}
    \centering
    \includegraphics[width=0.7\columnwidth]{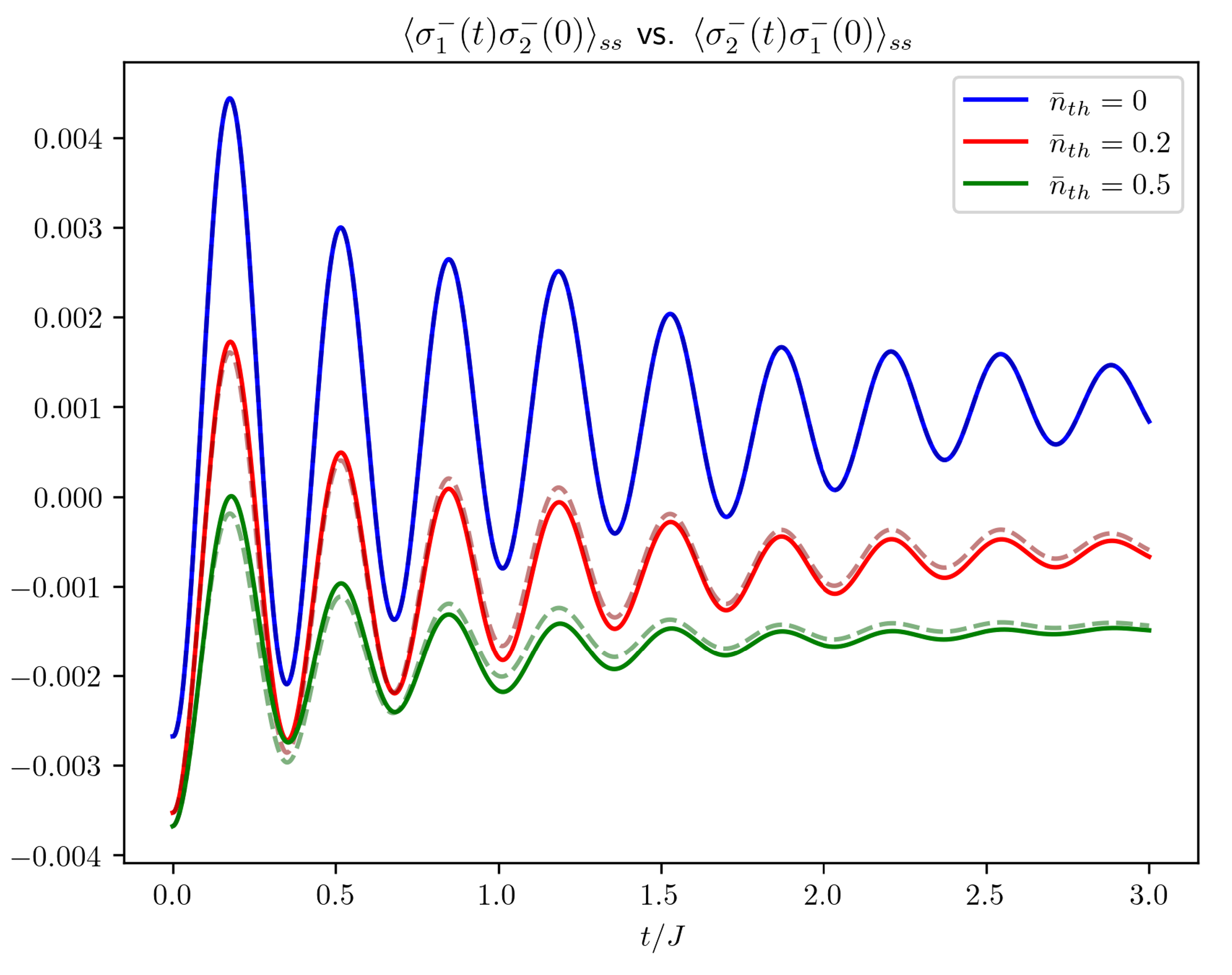}
    \caption{\textbf{Correlation function symmetry (and asymmetry) in the infinite-range dissipative transverse-field Ising model}. Here, we plot the steady-state correlation functions $\langle \hat{\sigma}^-_1(t)\hat{\sigma}^-_2 \rangle_\text{ss}$ (solid lines) and $\langle \hat{\sigma}^-_2(t)\hat{\sigma}^-_1(0) \rangle_\text{ss}$ (dashed lines), for different values of the temperature $\bar{n}_\text{th}$ (c.f. Eq. \eqref{eq:thermalmeq}). Here, $\kappa = J,h^z = J,N = 5$, and $h^x$ is sampled from a uniform distribution on $[0, 10J]$. In this plot, $h^x = \{9.2J, 2.2J, 5.2J, 4.4J, 1.0J\}$.}
    \label{fig:enter-label}
\end{figure}

\subsubsection*{Hidden TRS as a dynamical symmetry}
We now review the salient points of \cite{roberts_hidden_2020}. In particular, for any pair of operators $\hat{X},\hat{Y}$ that are left invariant under the exchange superoperator $\mathcal J$, the following correlation function symmetry holds:
\begin{align}
    \langle \hat{X}(t) \hat{Y}(0)\rangle_\text{ss} = \langle \hat{Y}(t) \hat{X}(0)\rangle_\text{ss}.
\end{align}
In our case, the detailed balance conditions \eqref{eq:the_constraint_Psi} imply that the effective Hamiltonian and jump operators are all left invariant under the exchange superoperator, and so, if we find a solution, then e.g. the following dynamical constraints hold in our model:
\begin{align}
        \forall i,j:~\langle \hat{\sigma}^-_i(t) \hat{\sigma}^-_j(0)\rangle_\text{ss} = \langle \hat{\sigma}^-_j(t) \hat{\sigma}^-_i(0)\rangle_\text{ss},\label{eq:symmetry}\\
        \forall i:~\langle \hat{\sigma}^-_i(t) \hat{H}_\text{eff}(0)\rangle_\text{ss} = \langle \hat{H}_\text{eff}(t) \hat{\sigma}^-_i(0)\rangle_\text{ss}
\end{align}
To illustrate this fact, we explicitly evaluate the steady-state correlation functions in \eqref{eq:symmetry} for a disordered transverse-field Ising model at finite temperature, which corresponds to the master equation
\begin{align}
    \partial_t \hat{\rho} = -i[\hat{H},\hat{\rho}] + \kappa (1+\bar{n}_\text{th})\sum_j\mathcal D[\hat{\sigma}_j^-] + \kappa \bar{n}_\text{th}\sum_j\mathcal D[\hat{\sigma}_j^+], \label{eq:thermalmeq}
\end{align}
where $\hat{H}$ is given in \eqref{eq:theham}.

\subsection{Solution in permutation-symmetric case}
We now assume the external magnetic field is homogeneous, $h^\mu_j \equiv h^\mu$. Then the constraint Eq. \eqref{eq:the_constraint_Psi} is permutation symmetric, and furthermore has at most one solution (otherwise the master equation would have multiple steady states, which is impossible, as was discussed in the previous section).  Therefore, any solution $\hat{\Psi}$, if it exists, must be permutation symmetric. Now suppose in Eq. \eqref{eq:the_constraint_Phi} that we choose a $\hat{K}$ which is permutation symmetric. Then $\hat{\Phi} = \hat{\Psi}\hat{K}$ is also permutation symmetric. If we further choose $\hat{K}$ so that the quadratic casimir for the collective $\mathfrak{su}(2)$ representation $\hat{S}^x,\hat{S}^y,\hat{S}^z$, i.e.
\begin{align}
    \hat{\mathfrak{C}}=(\hat{S}^x)^2 + (\hat{S}^y)^2 + (\hat{S}^z)^2 
\end{align}
commutes with $\hat{K}$, then any monomial $\hat{\mathfrak{C}}^m$ also commutes with $\hat{K}$, and therefore we can add any polynomial in the Casimir element with complex coefficients to any of the operators $\hat{O}$ in \eqref{eq:the_constraint_Phi}, without changing the space of solutions. In particular, this is useful for the case $\hat{O}=\hat{H}_\text{eff}$, which we can then deform into a non-Hermitian Ising model:
\begin{align}
    \hat{G}_\text{eff}&:=\hat{H}_\text{eff}+\frac{i\gamma N}{4} + \frac{i\Gamma \hat{\mathfrak{C}}}{2}= J_\text{eff}\hat{S}_z^2 + h^z_\text{eff}\hat{S}_z+ \frac{1}{2}(h\hat{S}_+ + h.c.),
\end{align}
where $J_\text{eff}=J+i\Gamma/2$, and we have introduced two effective complex magnetic fields: $h^z_\text{eff}=h^z-i(\gamma+\Gamma)/2$, and $h=h_x-ih_y$. Choosing the antiunitary $\hat{K} = \hat{K}_z\hat{U}_x$ referenced in the main text, we obtain:
\begin{align}
    (\hat{K}^\dagger \hat{G}_\text{eff} \hat{K})^\dagger=J_\text{eff}\hat{S}_z^2 - h^z_\text{eff}\hat{S}_z+ \frac{1}{2}(h\hat{S}_+ + h.c.),
\end{align}
Furthermore, the jump operators are $\hat{K}$-invariant: $(\hat{K}^\dagger \hat{\sigma}^-_j \hat{K})^\dagger = \hat{\sigma}^-_j.$ Therefore our condition Eq. \eqref{eq:the_constraint_Phi} reduces to $[\hat{\sigma}^-_j,\hat{\Phi}]=0,$ and
\begin{align}
    \hat{G}_\text{eff}\hat{\Phi} - \hat{\Phi} (\hat{K}^\dagger \hat{G}_\text{eff} \hat{K})^\dagger=J_\text{eff}[\hat{S}_z,\{\hat{S}_z,\hat{\Phi}\}] +h^z_\text{eff}\{\hat{S}_z,\hat{\Phi}\}  + \frac{1}{2}[(h\hat{S}_+ + h.c.),\hat{\Phi}] = 0.
\end{align}
Substituting $[\hat{S}_-,\hat{\Phi}]=0$ yields
\begin{align}
    \frac{\hat{G}_\text{eff}\hat{\Phi} - \hat{\Phi} (\hat{K}^\dagger \hat{G}_\text{eff} \hat{K})^\dagger}{J_\text{eff}}=[\hat{S}_z,\{\hat{S}_z,\hat{\Phi}\}] +r\{\hat{S}_z,\hat{\Phi}\}  + \frac{h}{2J_\text{eff}}[\hat{S}_+,\hat{\Phi}] = 0,\label{eq:K_eff_constraint}
\end{align}
where we have used the definition $r\equiv h^z_\text{eff}/J_\text{eff}$ used in the main text. The only solution to the set of constraints  $[\hat{\sigma}^-_j,\hat{\Phi}]=0$ is a polynomial in the lowering operators with complex coefficients, that is, $\hat{\Phi}\in \mathbb{C}[\hat{\sigma}^-_1,\dots, \hat{\sigma}_N^-]$. Because $\hat{\Phi}$ is permutation symmetric, this is a symmetric polynomial. By the fundamental theorem of symmetric polynomials, $\hat{\Phi}$ can further be expanded as a polynomial in the elementary symmetric polynomials. Because $\hat{\sigma}^-_j$ squares to zero, the elementary symmetric polynomials are each proportional to some power of $\hat{S}_-$. Therefore we can write
\begin{align}
    \hat{\Phi} =\sum_k c_k \hat{S}_-^k.
\end{align}
To solve for the coefficients $\{c_k\}_k$, we substitute the above polynomial into Eq. \eqref{eq:K_eff_constraint}. Defining $\hat{B}_k :=(2\hat{S}_z+k)\hat{S}_-^k$, and using standard spin commutation relations, we get
\begin{align}
    \{\hat{S}_z,\hat{S}_-^k\}= \hat{B}_k,~~~[\hat{S}_+,\hat{S}_-^k]= k\hat{B}_{k-1},~~~[\hat{S}_z, \hat{B}_k] = -k\hat{B}_k,
\end{align}
so that, noting that $\hat{B}_N = 0$, we get
\begin{align}
    \frac{\hat{G}_\text{eff}\hat{\Phi} - \hat{\Phi} (\hat{K}^\dagger \hat{G}_\text{eff} \hat{K})^\dagger}{J_\text{eff}}= \sum_{k=0}^{N-1}\bigg(\frac{h}{2J_\text{eff}}(k+1)c_{k+1}-(k-r)c_k\bigg)\Phi_k=0.
\end{align}
Since the operators $\hat{B}_k$ are linearly independent, each coefficient must vanish, yielding a first-order recursion relation for $\{c_k\}_k$, which yields the binomial series solution
\begin{align}
    \hat{\Phi} = \sum_{k=0}^\infty \binom{r}{k}(-\hat{\mathbb{S}}_-)^k,
\end{align}
where $\hat{\mathbb{S}}_-$ is the non-unitary $SU(2)$ lowering operator defined in the main text, and $\binom{z}{k} = z(z-1)\cdots (z-k+1)/k!$ is the generalized binomial coefficient. Explicitly,
\begin{align}
    \hat{\mathbb{S}}_- = \frac{2J_\text{eff}}{h}\hat{S}_-.
\end{align}
In the following section, we will show how one can generalize this solution to the case where the transverse-plane magnetic fields $h_j:= h^x_j-ih_j^y$ have arbitrary spatial dependence, but at the cost of removing the collective loss term \eqref{eq:collectiveLossTerm} from the Lindbladian.

\subsection{Solution without permutation symmetry, $\Gamma = 0$}

We now consider the following conditions but possibly allowing for a spatially-inhomogeneous magnetic field:
\begin{align}
    \hat{O}\hat{\Phi}=\hat{\Phi}(\hat{K}^\dagger \hat{O} \hat{K})^\dagger~~~~~~\forall \hat{O}\in \text{span}\{\hat{H}_\text{eff}, \hat{L}_1, \cdots, \hat{L}_l\}\label{eq:A1}
\end{align}
In this case it seems that the lack of permutation symmetry would invalidate all of the arguments we used previously to calculate the solution $\hat{\Phi}$. Indeed, we cannot add a polynomial in the Casimir element to $\hat{H}_\text{eff}$ to obtain $\hat{G}_\text{eff}$, and are instead stuck with an inhomogeneous mixed-field XXZ model:
\begin{align}
    \hat{H}_\text{eff}= J\hat{S}_z^2-\frac{i\Gamma}{2}(\hat{S}_x^2+\hat{S}_y^2) + h^z_\text{eff}\hat{S}_z+ \frac{1}{2}\sum_{j}(h_j\hat{\sigma}^+_j + h^*_j\hat{\sigma}_j^-)
\end{align}
However, we can still solve the detailed balance condition if $\Gamma=0$. In this case the imaginary XY term vanishes. Using the same choice of $\hat{K}$ as before, the condition Eq. \eqref{eq:the_constraint_Phi} with $\hat{O} = \hat{H}_\text{eff}$ reduces to
\begin{align}
    \frac{\hat{H}_\text{eff}\hat{\Phi} - \hat{\Phi} (\hat{K}^\dagger \hat{H}_\text{eff} \hat{K})^\dagger}{J}=[\hat{S}_z,\{\hat{S}_z,\hat{\Phi}\}] +r\{\hat{S}_z,\hat{\Phi}\}  + \sum_j\frac{h_j}{2J}[\hat{\sigma}^+_j,\hat{\Phi}] = 0.
\end{align}
The inhomogeneity in the transverse-plane magnetic field can now be removed via a similarity transformation $\hat{\mathcal G}:=\prod_je^{\hat{S}_j^z\log (2J_\text{eff}/h_j)}$, which reduces the above condition to a special case of Eq. \eqref{eq:K_eff_constraint}:
\begin{align}
    \hat{\mathcal G}\frac{\hat{H}_\text{eff}\hat{\Phi} - \hat{\Phi} (\hat{K}^\dagger \hat{H}_\text{eff} \hat{K})^\dagger}{J}\hat{\mathcal G}^{-1}=[\hat{S}_z,\{\hat{S}_z,\hat{\Phi}_{\mathcal G}\}] +r\{S_z,\hat{\Phi}_{\mathcal G}\}  + [\hat{S}_+,\hat{\Phi}_{\mathcal G}] = 0,\label{eq:inhom_constraint}
\end{align}
where $\hat{\Phi}_{\mathcal G} \equiv \hat{\mathcal{G}}^{-1}\hat{\Phi} \hat{\mathcal G}$ is the transformed solution. Comparing Eq. \eqref{eq:inhom_constraint} with Eq. \eqref{eq:K_eff_constraint} immediately yields
\begin{align}
    \hat{\Phi}_{\mathcal G} = (1-\hat{S}_-)^r.
\end{align}
The general solution cited in the main text then follows from the fact that our two $SU(2)$ representations are precisely related by this similarity transformation: $\hat{\mathbb{S}}^\mu = \hat{\mathcal G}\hat{S}^\mu \hat{\mathcal G}^{-1}$.


\section{Spin blockade}

We now derive the bounds used in the main text to determine the threshold where blockade effects become observable. A good estimate for this threshold is obtained as the point where $p_n = p_N$, where $\{p_m\}_m$ is the probability distribution defined in the main text. We now calculate this threshold:
\begin{align}
    p_n = \bigg\|\binom{r}{n}\hat{\mathbb{S}}_-^n \bigg\|_\text{HS}^2,~~~~~~~p_N =  \bigg\|\binom{r}{N}\hat{\mathbb{S}}_-^N\bigg\|_\text{HS}^2
\end{align}
We calculate
\begin{align}
    \frac{p_N}{p_n}&= \frac{n!^2}{N!^2}|(-\epsilon)(-\epsilon-1)\cdots (-\epsilon - N+n+1)|^2\frac{\|\hat{\mathbb{S}}_-^N\|_\text{HS}^2}{\|\hat{\mathbb{S}}_-^n\|_\text{HS}^2}.
\end{align}
We calculate the $L^2$ norm of $\hat{\mathbb{S}}_-^n$ by brute force using the multinomial expansion:
\begin{align}
    \|\hat{\mathbb{S}}_-^n\|_\text{HS}^2 = n!^2\sum_{X\subset \{1,2,\cdots, N\},~|X|=n} \bigg\|\prod_{j\in X}\frac{2J_\text{eff}}{h^x_j}\hat{\sigma}_j^-\bigg\|^2_\text{HS}=n!^22^N\mathcal I_n,
\end{align}
where $\mathcal I_n \equiv \mathcal I_n(\{1,2,\dots, N\})$ is the form factor defined in the main text. We thus obtain the asymptotic estimate
\begin{align}
    \frac{p_N}{p_n} &\sim |\epsilon|^2\times (N-n-1)!^2 \frac{\mathcal I_n}{\mathcal I_N}~~~~~(\epsilon\to 0)
\end{align}
Setting the ratio equal to one yields the result $|\epsilon| <(N-n-1)!^{-1}\sqrt{\mathcal I_n/\mathcal I_N}$. In the uniform driving limit, $\mathcal I_n = \lambda^n \binom{N}{n}$, where $\lambda = 2|J_\text{eff}/h^x|^2$ is the dimensionless driving parameter defined in the main text. This yields the bound
\begin{align}
    |\epsilon|<\frac{1}{(N-n-1)!}\sqrt{\lambda^{n-N}\binom{N}{n}}.
\end{align}

\subsubsection{Blockade effect with $\gamma \approx J$}
\begin{figure}
    \centering
    \includegraphics[width=0.55\columnwidth]{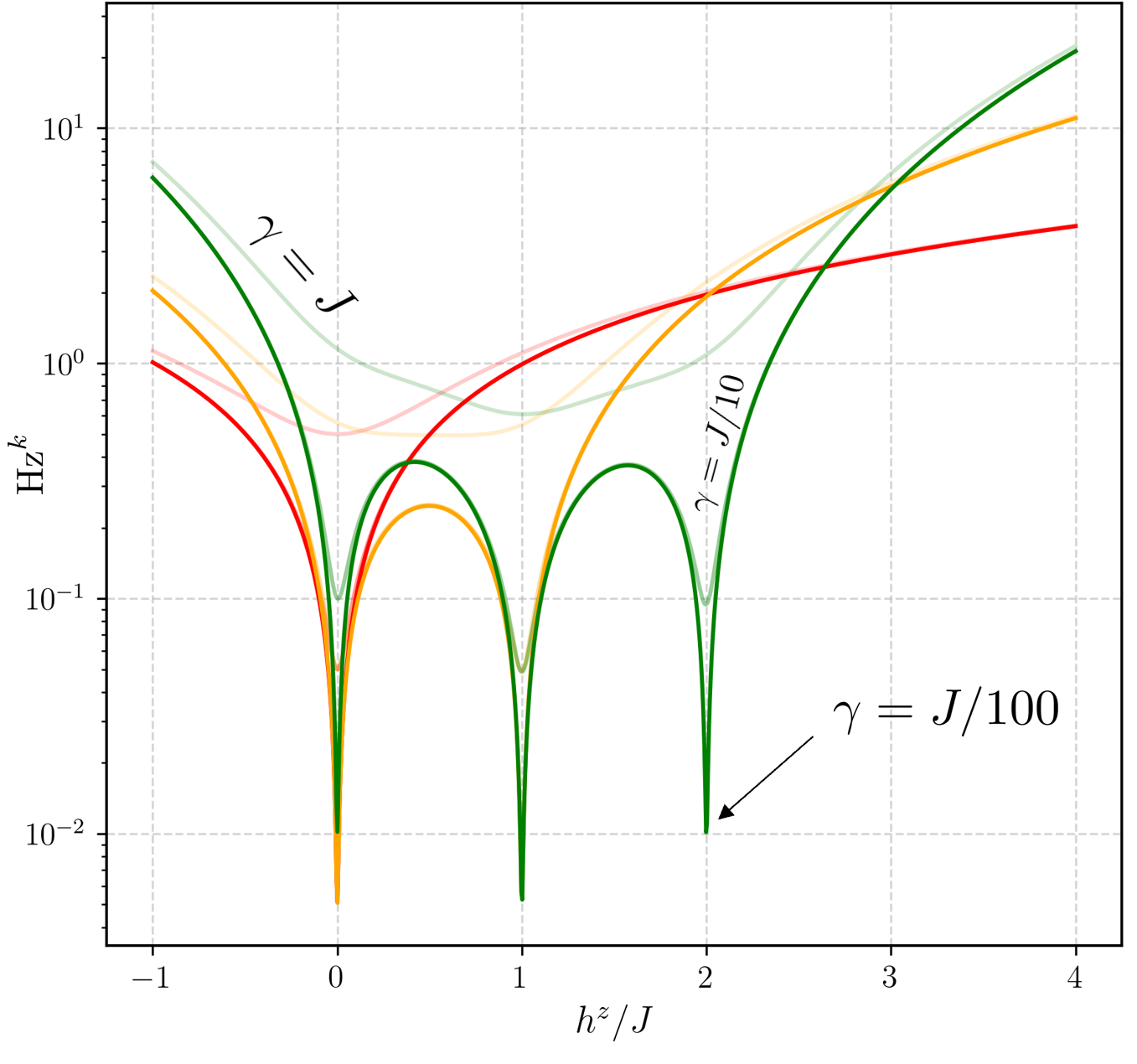}
    \caption{Spin blockade effect with $\|\hat{\mathbb{S}}_-\|_\text{HS}=2/3$. Rescaled coherence tensor $h_{i_1}\cdots h_{i_k}C_{i_1,\cdots, i_k}$, for $k=1$ (red), $k=2$ (orange) and $k=3$ (green). Here, $N=22,h^x = 3NJ,h^y =0$, and $\Gamma = 0$. $\gamma \in \{J,J/10,J/100\}$, with greater transparency indicating greater loss rates. Note how the dips in the rescaled coherence are noticeable even with loss rates $\gamma$ comparable to the interaction strength $J$. Furthermore, these dips have a sizeable width $\sim O(1)$.}
    \label{fig:blockade_extra}
\end{figure}

Despite what the previous discussion might suggest, the condition $\gamma \ll J$ is not strictly necessary for observing the blockade effect. For this purpose, let us assume that the Hilbert-Schmidt norm $\|\hat{\mathbb{S}}_-\|_\text{HS}$ is less than one half. This effectively  sets $h_j/J_\text{eff}\sim O(N)$, or equivalently $h_j/\bar{J}_\text{eff}\sim O(1)$. In this case,
\begin{align}
    \|\log (1-\hat{\mathbb{S}}_-)\|_\text{HS}\leq \log (1-\|\hat{\mathbb{S}}_-\|_\text{HS}) <1.
\end{align}
As a result, the following series uniformly converges when $|\epsilon|<1$:
\begin{align}
    \hat{\Phi} &=(1-\hat{\mathbb{S}}_-)^n\bigg(1 +  \epsilon\log (1-\hat{\mathbb{S}}_-)+ \cdots \bigg).
\end{align}
In this case, the threshold to observe an effective blockade is $|\epsilon| < 1$, and thus the effect remains observable well into the thermodynamic limit. However, the constraint $\|\mathbb{S}_-\|_\text{HS}<1/2$ means that the steady state is close to an infinite-temperature state. Representative results are shown in Figure \ref{fig:blockade_extra}.

\subsection{Distribution for the magnetization}
We now calculate the distribution for the total magnetization at a blockade. We attempt to calculate the quantity,
\begin{align}
    p(q_1,\cdots, q_N) = \langle q_1,\cdots, q_N |\hat{\Phi}\hat{\Phi}^\dagger |q_1,\cdots, q_N\rangle = \|\hat{\Phi}^\dagger |q_1,\cdots, q_N\rangle\|^2
\end{align}
where $q_j$ is a Boolean variable that is zero if the $j$th spin is pointing up, and one otherwise. In the limit $h^x_j/J_\text{eff}\ll 1$ we can approximate $\hat{\Phi}\approx \hat{\mathbb{S}}_-^n$, and we obtain
\begin{align}
    p(q_1,\cdots, q_N) = \langle q_1,\cdots, q_N |\hat{\Phi}\hat{\Phi}^\dagger |q_1,\cdots, q_N\rangle = \|\hat{\Phi}^\dagger |q_1,\cdots, q_N\rangle\|^2 \sim \sum_{A\subset \{1,2,\dots, N\},~|A|=n}\prod_{j\in A}\lambda_j q_j,
\end{align}
where $\lambda_j = 2|J_\text{eff}/h^x_j|^2$. To obtain the distribution for the total magnetization, it suffices to coarse-grain the above distribution over $k = \sum_j q_j$. We obtain
\begin{align}
    p_k = \sum_{\sum_j q_j = k} p(q_1,\cdots, q_N) &= \sum_{B\subset \{1,2,\dots, N\},~|B|=k} \sum_{A\subset B,~|A|=n} \prod_{j\in A}\lambda_j= \sum_{A\subset \{1,2,\cdots, N\},~|A|=n}\Bigg(\prod_{j\in A}\lambda_j \sum_{B\supset A,~|B|=k} 1\Bigg)\\
    &= \sum_{A\subset \{1,2,\cdots, N\},~|A|=n}\Bigg(\prod_{j\in A}\lambda_j \binom{N- n}{k-n}\Bigg) = \binom{N- n}{k-n}\mathcal I_n \propto \binom{N}{k} \binom{k}{n},
\end{align}
which yields the binomial distribution given in the main text.

\section{Thermodynamic limit $N\to\infty$}

We now calculate the distribution $\{p_M\}_M$ in the limit that $N\to\infty$. To make this task easy, we assume that the external magnetic fields are uniform, $h^\mu_j \equiv h^\mu$. The exact expression is

\begin{align}
    p_M = \bigg|\binom{r}{M}\bigg|^2\|\hat{\mathbb{S}}_-^M\|_\text{HS}^2=&2^{N-M} M!^2 |2J_\text{eff}/h|^{2M} \binom{N}{M}\bigg|\binom{r}{M}\bigg|^2=|(-r)_M|^2\frac{N!}{(N-M)!} \frac{\lambda^M}{M!}\nonumber\\
    &=|(-N\bar{r})_M|^2 \frac{N!}{(N-M)!}\frac{\bar{\lambda}^M}{N^{2M}M!},\label{eq:largeNeq1}
\end{align}
where we have dropped all $M$-independent constants, and have recalled the definitions in the main text:
\begin{align}
    \bar{r}\equiv \frac{h^z_\text{eff}}{\bar{J}_\text{eff}},~~~~~~~~\bar{\lambda} =  2|\bar{J}_\text{eff}/h^x|^2,
\end{align}
where $\bar{J}_\text{eff}\equiv NJ_\text{eff}$. We now expand 
\begin{align}
    (-N\bar{r})_M& = N^M(-\bar{r})\cdot \bigg(\frac{1}{N}- \bar{r}\bigg)\cdots \bigg(\frac{M-1}{N}- \bar{r}\bigg),~~~~~~    \frac{N!}{(N-M)!} = N^M1\cdot \bigg(1-\frac{1}{N}\bigg)\cdots \bigg(1-\frac{M-1}{N}\bigg).
\end{align}
Substituting $m\equiv M/N$ allows us to write the logarithm of \eqref{eq:largeNeq1} as a Riemann sum:
\begin{align}
    N^{-1}\log p_M=m\log N\bar{\lambda} - N^{-1}\log M!+\sum_{m'\in \{k/N\}_{k=0}^{M-1}}\log (1-m')|\bar{r}-m'|^2\Delta m' 
\end{align}
with $\Delta m' \equiv 1/N$. We can simplify the terms excluding the Riemann sum by using Stirling's approximation $\log M! = M\log M - M + O(\log M)$:
\begin{align}
    m\log N\bar{\lambda} - N^{-1}\log M!=m\bigg(1+\log \frac{\bar{\lambda}}{m}\bigg)  + O(N^{-1}),
\end{align}
where again we have neglected $M$-independent constants. We therefore arrive at the asymptotic estimate
\begin{align}
    f(m) =-\int_0^mdm'\log (1-m')|\bar{r}-m'|^2 - m\bigg(1+\log \frac{\bar{\lambda}}{m}\bigg) + O(N^{-1}).
\end{align}
Carrying out the integral analytically yields the closed-form expression utilized in the main text.

\subsection{$1/N$ corrections to the potential}

The leading-order corrections to the potential come from both (1) including the $O(N^{-1})$ terms in Stirling's approximation, and (2) writing the Riemann sum as an asymptotic series (whose terms are each Riemman integrals). For (1), note the improved estimate
\begin{align}
    m\log N\bar{\lambda} - N^{-1}\log M!=m\bigg(1+\log \frac{\bar{\lambda}}{m}\bigg) +\frac{1}{2N}\log m + O(N^{-2}),
\end{align}
where again, we have dropped $M$-independent constants. Finally, for the Riemann sum, we note that the integrand $g(m) \equiv \log (1-m)|\bar{r}-m|^2$ is a real-analytic function on the domain $m\in (0,1)$. This is because $r$, as was discussed in the main text, cannot be a non-negative real number. Therefore, we have the asymptotic series expansion
\begin{align}
    \int_0^mdm'\,g(m')= \sum_{m'\in \{k/N\}_{k=0}^{M-1}} &g(m')\Delta m' + \sum_{k=1}^\infty \frac{(\Delta m')^m}{(k+1)!}\sum_{m'\in \{k/N\}_{k=0}^{M-1}}g^{(k)}(m')\Delta m',\\
    &= \sum_{m'\in \{k/N\}_{k=0}^{M-1}} g(m')\Delta m' + \frac{\Delta m'}{2}\int_0^m dm'\,g'(m')  + O(N^{-2}).\nonumber
\end{align}
This leads to the following improved estimate for the Riemann sum:
\begin{align}
    \sum_{m'\in \{k/N\}_{k=0}^{M-1}}\log (1-m')|\bar{r}-m'|^2\Delta m'  &= \int_0^mdm'\, \log (1-m')|\bar{r}-m'|^2\nonumber \\
    &- \frac{1}{2N}\log (1-m)|\bar{r}-m|^2 + O(N^{-2})
\end{align}
Putting both $O(N^{-1})$ corrections together yields the following shift in the free energy:
\begin{align}
    \delta f(m) = \frac{1}{2N}\log \frac{|m-\bar{r}|^2(1-m)}{m} + O(N^{-2}).\\
    \nonumber
\end{align}

\subsection{Critical points of the potential}
To understand the large-$N$ behavior of our infinite-range DTI model, it is useful to calculate the derivative of the potential:
\begin{align}
    f'(m) = \log \frac{m}{\bar{\lambda}(1-m)|m-\bar{r}|^2}.
\end{align}
The critical equation $f'(m)=0$ is thus equivalent to the cubic equation $m = \bar{\lambda}(1-m)|m-\bar{r}|^2$. On the other hand, the mean-field equations for $S_\mu := \langle \hat{\sigma}^\mu\rangle_\text{SS}$ are
\begin{align}
    0&= -2\bar{J}S_yS_z-\frac{\tilde{\gamma}}{2}S_x+\bar{\Gamma} S_x S_z + i(h^zS_y - h^y S_z),\label{eq:mx}\\
    0 &= -2\bar{J}S_xS_z-\frac{\tilde{\gamma}}{2}S_y+\bar{\Gamma} S_y S_z + ih^x+ i(h^xS_z - h^z S_x),\label{eq:my}\\
    0 &= -\frac{\tilde{\gamma}}{2}(2S_z+1)-\bar{\Gamma}(S_x^2 + S_y^2) +i(h^yS_x - h^xS_y),\label{eq:mz}
\end{align}
where $\tilde{\gamma} = \gamma + \bar{\Gamma}/N$, with $\bar{\Gamma}:= N\Gamma$ \cite{huybrechtsMeanfieldValidityDissipative2020}. Writing $S_\mu = 2m_\mu$, and substituting Eqs. \eqref{eq:mx}, \eqref{eq:my} into Eq. \eqref{eq:mz}, while letting $N\to\infty$, we obtain
\begin{align}
    m_z^3 +(1+2\text{Re}\, \bar{r}) m_z^2 +|\bar{r}|^2 + (|\bar{r}+1|^2 -1+\bar{\lambda}^{-1})m_z=0,
\end{align}
which is precisely the critical equation $f'(m) = 0$, once we substitute the identity $m = -m^z$. This demonstrates that the critical points of $f(m)$ are in one-to-one correspondence with the mean-field solutions for $m_z$, for {\it all} possible values of $\gamma,\Gamma,J,h^x,h^y,h^z$ (forming a 6-dimensional parameter space). 

\newpage

\section{Correlation functions}
We now proceed to calculate the expectation values of normally-ordered strings of Pauli operators. This process largely adheres to the formalism described in the supplementary information (SI) of \cite{roberts_competition_2023}. It's worth noting that the particular methodology of \cite{roberts_competition_2023} was in fact initially devised during the investigation of the current model, for which the calculation is much simpler:
\begin{align}
    \hat{\sigma}^+_{C_1}\hat{\sigma}^-_{C_2}\equiv \prod_{i\in C_1} \hat{\sigma}^+_i\prod_{j\in C_2}\hat{\sigma}^-_j,
\end{align} 
where $C_1,C_2\subset\{1,2,3,\dots, N\}$ are the clusters of spins mentioned in the main text. We use the multinomial expansion $\hat{\mathbb{S}}_-^k= k!\sum_{|A|=k}\alpha^A \hat{\sigma}^-_A$ to write 
\begin{align}
    \langle \hat{\sigma}^+_{C_1}\hat{\sigma}^-_{C_2}\rangle_\text{SS}&= \frac{1}{\mathcal N}\sum_{k,l}(-1)^{k+l}\binom{r^*}{k}\binom{r}{l}\text{tr}(\hat{\mathbb{S}}_-^{\dagger k}\hat{\sigma}^+_{C_1}\hat{\sigma}^-_{C_2}\hat{\mathbb{S}}_-^l)\nonumber\\
    &=\frac{1}{\mathcal N}\sum_{A_1,A_2\subset \{1,2,3,\dots, N\}}(-r^*)_{|A_1|}(-r)_{|A_2|} \alpha^{*A_1} \alpha^{A_2} \text{tr}(\hat{\sigma}^+_{A_1}\hat{\sigma}^+_{C_1}\hat{\sigma}^-_{C_2}\hat{\sigma}^-_{A_2}).\label{eq:big_sum}
\end{align}
where $\alpha_j \equiv 2J_\text{eff}/h_j$, and we have used the shorthand notation $\alpha^A\equiv \prod_{j\in A}\alpha_j$. The fundamental trace $\text{tr}(\hat{\sigma}^+_{A_1}\hat{\sigma}^+_{C_1}\hat{\sigma}^-_{C_2}\hat{\sigma}^-_{A_2})$ inside the sum vanishes unless $A_j\subseteq C_j^c$, and $A_1 \amalg C_1= A_2 \amalg C_2$. Because of the former constraint, we can write
\begin{align}
    A_j = A_j\cap R_{12}^c \,\amalg A_j\cap R_j\\
    \nonumber
\end{align}
where $\amalg$ denotes the disjoint union, and $R_{12},R_j$ are the regions defined in the main text. The constraint $A_1 \amalg C_1= A_2 \amalg C_2$, yields
\begin{align}
    ~~A_j\cap R_j = R_j,~~~~~~~A_1\cap R_{12}^c &= A_2\cap R_{12}^c \equiv X.\\
    \nonumber
\end{align}
Furthermore, we can write $A_j = X\amalg R_j$ and $A\amalg C_j = X\amalg R_{12}$, and hence $(A_1,A_2)\to X \equiv  A_j\setminus R_j$ is a bijection from the support of the sum in Eq. \eqref{eq:big_sum}, to the power set (the set of all subsets) of $R_{12}^{c}$. Therefore,
\begin{align}
    \langle \hat{\sigma}^+_{C_1}\hat{\sigma}^-_{C_2}\rangle_\text{SS}&=\frac{\alpha^{*R_1}\alpha^{R_2} (-r^*)_{|R_1|}(-r)_{|R_2|}}{\mathcal N}\nonumber\\
    &~~~~~~~~~~~~~~~~~~~~~\times\sum_{X\subset R_{12}^c} (|R_1|-r^*)_{|X|}(|R_2|-r)_{|X|} 2^{N-|X|-|R_{12}|} \prod_{j\in X} |\alpha_j|^2,\label{eq:general_soln_correlations}
\end{align}
Eq. \eqref{eq:general_soln_correlations} immediately leads to the formula given in the main text.\\

\begin{figure}
    \centering
    \includegraphics[width=0.5\columnwidth]{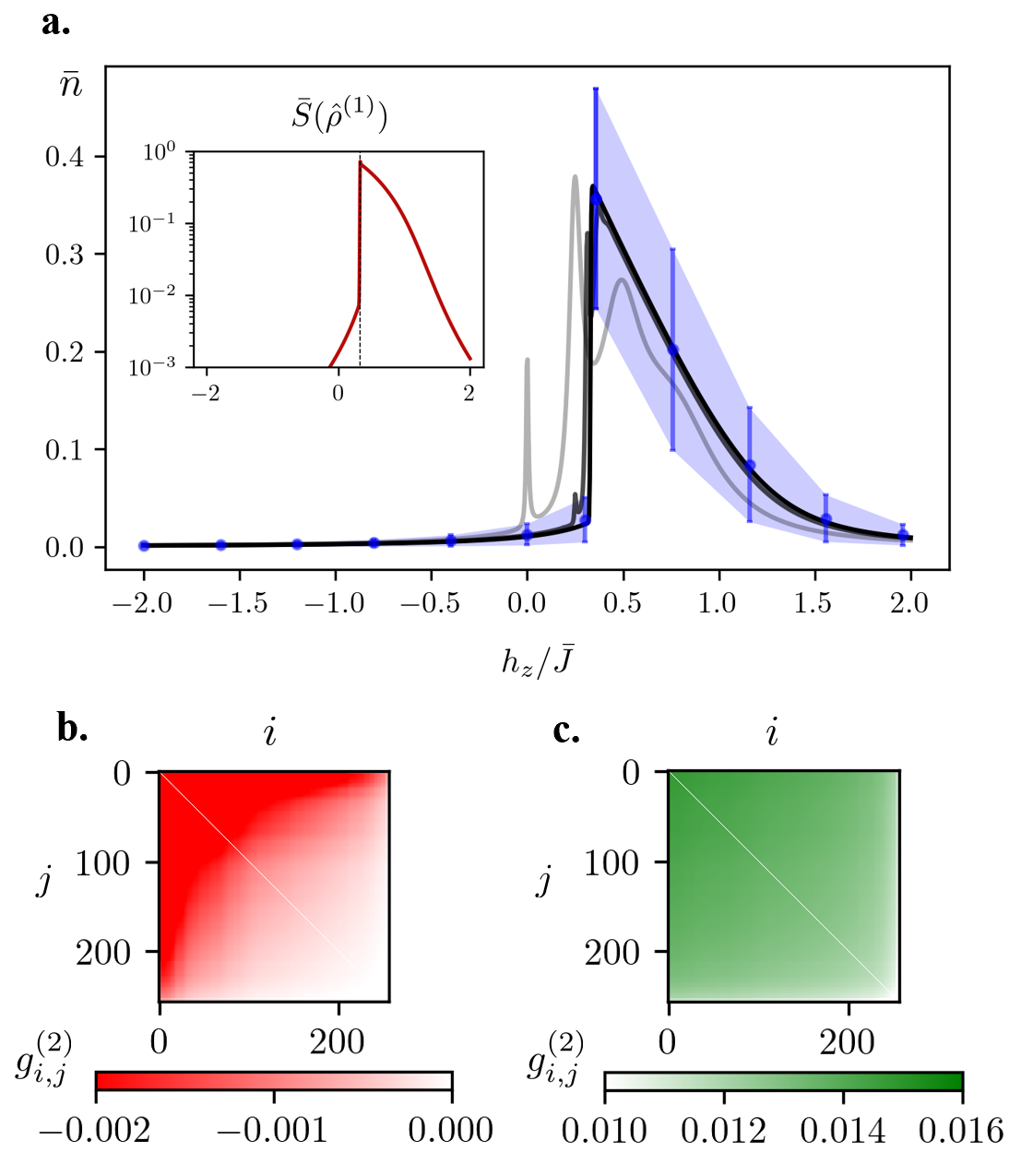}
    \caption{{\bf Dissipative phase transitions in a random transverse-field Ising model}. (a) Solid curves: density of up spins $\bar{n} = \sum_j \langle \hat{\sigma}^+_j\hat{\sigma}^-_j\rangle_\text{SS}/N$, for $N\in 4$ (lightest gray), $N=16$ (light gray), and $N = 128$ (black). Parameters are $\gamma = \bar{J}/100$, $\Gamma = 0$, and $h^x = \bar{J}/5, h^y = 0$. Blue dots and error bars denote $\bar{n}$ and $\Delta\bar{n}\equiv \sqrt{\sum_j (\langle \hat{\sigma}^+_j\hat{\sigma}^-_j\rangle_\text{SS} - \bar{n})^2}$ respectively, for a $N=128$ random transverse-field Ising model with the same parameters, but with $h_j^x$ sampled from a normal distribution with mean $h^x$ and variance $\sigma=h^x/2$.  Results are all for a single disorder realization. Inset: von-Neumann entropy of the 1-spin reduced density matrix for the non-random transverse-field Ising model with $N=128$ (black curve). Bottom panel: connected density-density correlation function $g^{(2)}_{i,j}$ for a $N=256$ random transverse-field Ising model with the same parameters (i.e. $h_j^{x,y}, \gamma, \Gamma, J$) as in panel (a), but with $h^z =0.30\bar{J}$ (b) and $h^z=0.36\bar{J}$ (c). }
    \label{fig:fancy_disorder_figure}
\end{figure}

\subsection{Permutation-symmetric limit}

With permutation symmetry, the recursion relation for $\mathcal I_m(X)$ presented in the main text admits the closed-form solution 
\begin{align}
    \mathcal I_m(X) = \binom{|X|}{m}(2|J_\text{eff}/h|^2)^m.
\end{align}
Substituting this solution for $\mathcal I_m$ into the general formula in the main text yields
\begin{align}
    \langle \hat{\sigma}^+_{C_1}\hat{\sigma}^-_{C_2}\rangle_\text{SS}&=\frac{2^{|R_{12}^c|}\mathcal K^*(R_1)\mathcal K(R_2)}{\mathcal N}\sum_m (|R_1|-r^*)_m(|R_2|-r)_m\frac{(-|R_{12}^c|)_m}{m!}\bigg(-\frac{2|J_\text{eff}|^2}{|h_j|^2}\bigg)^m\nonumber
\end{align}
which leads to the solution used in the main text, once we recall the definition of the generalized hypergeometric function:
\begin{align}
    \,_pF_q(a_1,\dots a_p;b_1,\dots b_q;z)= \sum_{l=0}^\infty \frac{\prod_{m=1}^p(a_m)_l}{\prod_{m=1}^q(b_m)_l}\frac{z^l}{l!}.\\
    \nonumber
\end{align}


\subsection{Phase transitions in the disordered model}
We use the above solutions for correlation functions to investigate phase transitions in the disordered version of our model. We plot these correlation functions for single disorder realizations in Figure \ref{fig:fancy_disorder_figure}, where we obtain the following preliminary findings:
\begin{itemize}
    \item The phase transition remains sharp, even with strong disorder. The location of the transition also doesn't seem to shift at all, even with transverse-field disorder $\Delta h^x/h^x\sim O(1)$.
    \item As in \cite{roberts_competition_2023}, the sign of the connected density-density correlation function
    \begin{align}
        g^{(2)}_{i,j} \equiv \frac{\langle \hat{n}_i\hat{n}_j\rangle-\bar{n}_i\bar{n}_j}{\bar{n}_i\bar{n}_j}
    \end{align}
    flips at the transition point. Note: in panels (b) and (c) of Figure \ref{fig:fancy_disorder_figure}, the spins are reordered so that $h_j^x$ is an increasing function of $j$.
\end{itemize}

\bibliography{SI-refs}